\documentclass[aps,12pt,nofootinbib,showpacs,showkeys,preprintnumbers,amsmath,amssymb]{revtex4-1}

\usepackage{amsmath,amsfonts,amssymb,amsthm,bm,bbm,bbold,braket,color,dsfont,fancybox,hyperref,
srcltx,subfigure,ucs,yfonts,cancel,xfrac,verbatim,xcolor}
        
\usepackage{float}

\usepackage{lineno}
\usepackage{mathtools}
\newcommand{\beq}{\begin{equation}}
\newcommand{\eeq}{\end{equation}}
\newcommand{\bea}{\begin{eqnarray}}
\newcommand{\eea}{\end{eqnarray}}
\newcommand{\beas}{\begin{eqnarray*}}
\newcommand{\eeas}{\end{eqnarray*}}
\newcommand{\bi}{\begin{itemize}}
\newcommand{\ei}{\end{itemize}}

\usepackage{amsmath,amsfonts,amssymb,amsthm,bm,bbm,bbold,braket,color,dsfont,fancybox,hyperref,
srcltx,slashed,ucs,yfonts,cancel,xfrac,verbatim,xcolor}
\usepackage[inline,shortlabels]{enumitem}
\usepackage{nicefrac}
\usepackage{multirow}

\DeclareMathAlphabet{\mathpzc}{OT1}{pzc}{m}{it}
\definecolor{gold}{rgb}{1,0.8,0}
\definecolor{nara}{rgb}{1,0.4,0.1}
\definecolor{goldo}{rgb}{1,0.7,0}
\definecolor{greeno}{rgb}{0,0.8,0}
\def\bes{\begin{subequations}}
\def\ees{\end{subequations}}
\def\be{\begin{equation}}
\def\ee{\end{equation}}
\def\bea{\begin{eqnarray}}
\def\eea{\end{eqnarray}}
\def\ba{\begin{eqnarray}}
\def\ea{\end{eqnarray}}
\def\bear{\begin{array}}
\def\eear{\end{array}}

\newcommand{\bpm}{\begin{pmatrix}}
\newcommand{\epm}{\end{pmatrix}}
\newcommand{\BM}{\left(\begin{array}}		% matrices
\newcommand{\BMC}{\left[\begin{array}}		% matrices
\newcommand{\EM}{\end{array}\right)}		% matrices
\newcommand{\EMC}{\end{array}\right]}		% matrices
	% ...
	% ...
 % ...

\newcommand{\com}[1]{\begin{comment}\end{comment}}
\newcommand{\K}{\mathcal{K}}

\begin{document}

%\begin{flushright}
%\preprint{\textbf{OU-HEP-1014}}
%\end{flushright}

\title{Exploring CP-Violating heavy neutrino oscillations in rare tau decays at Belle II}

\author{Sebastian Tapia$^{1}$}
\email{s.tapia@cern.ch}
\author{Jilberto Zamora-Sa\'a$^{2}$}
\email{jilberto.zamora@unab.cl}
%\pacs{14.60.St,13.20-v,13.15.+g}

 \affiliation{$^1$Department of Physics, University of Illinois at Urbana-Champaign, Urbana, IL 61801, USA.}
 \affiliation{$^2$Departamento de Ciencias F\'isicas, Universidad Andres Bello,  Sazi\'e 2212, Piso 7,  Santiago, Chile.}
 %=================================================================================================================
 %=================================================================================================================
 %=================================================================================================================
 %=================================================================================================================
\begin{abstract}

In this work, we study the lepton number violating tau decays via two intermediate on-shell Majorana neutrinos $N_j$ into two charged pions and a charged lepton \\ $\tau^{\pm} \to \pi^{\pm} N_j \to \pi^{\pm} \pi^{\pm} \ell^{\mp}$. We consider the scenario where the heavy neutrino masses are within $0.5$ GeV $\leq M_N \leq 1.5$ GeV. We evaluated the possibility to measure the modulation of the decay width along the detector length for these processes at tau factories, such as Belle II. We study some realistic conditions which could lead to the observation of this phenomenon at futures $\tau$ factories.

\end{abstract}

\keywords{Heavy Neutrinos, Lepton Number Violation, Tau Factory, Heavy Neutrino Oscillations, Belle II.}

\maketitle

%=================================================================================================================
%=================================================================================================================
%=================================================================================================================
%=================================================================================================================
\section{Introduction}
\label{s1}
%=================================================================================================================
The first indications of physics beyond the standard model (SM) come from neutrino oscillations (NOs), 
baryonic asymmetry of the universe (BAU) and dark matter (DM). In the recent years NOs experiments have confirmed that 
active neutrinos ($\nu$) are very light massive particles $M_{\nu} \sim 1$ eV~\cite{Fukuda:1998mi,Eguchi:2002dm} and, consequently,
that the Standard Model must be extended. One of the most popular SM extensions, which explains very small neutrino masses, 
among others unknowns, is the See-Saw Mechanism (SSM) \cite{Mohapatra:2005wg,Mohapatra:2006gs}. The SSM introduces a new Majorana 
particle (SM-singlet) called heavy neutrinos (HN), which induces a dimension-5 operator \cite{Weinberg:1979sa} and leads to a very 
light active Majorana neutrino. Due to the fact that heavy neutrinos are singlet under the $SU(2)_L$  symmetry group, their 
interaction with gauge bosons ($Z,W^{\pm}$) and other leptons ($e, \mu, \tau$) must be highly suppressed. Despite this 
suppression, they can be searched via colliders~\cite{Das:2018usr,Das:2017nvm,Das:2012ze,Antusch:2017ebe,Das:2017rsu,
Das:2017zjc,Chakraborty:2018khw,Cvetic:2019shl,Antusch:2016ejd,Cottin:2018nms,Duarte:2018kiv,Drewes:2019fou,Dev:2019rxh,
Cvetic:2018elt,Cvetic:2019rms,Das:2018hph,Das:2016hof,Das:2017hmg}, rare meson decays~\cite{Dib:2000wm,Cvetic:2012hd,Cvetic:2013eza,Cvetic:2014nla,Cvetic:2015naa,
Cvetic:2015ura,Dib:2014pga,Zamora-Saa:2016qlk,Milanes:2018aku,Mejia-Guisao:2017gqp} and tau factories \cite{Zamora-Saa:2016ito,
Kim:2017pra,Dib:2019tuj}. Among the well-know SM extensions based on SSM, we can mention the Neutrino-Minimal-Standard-Model, 
$\nu$MSM~\cite{Asaka:2005an,Asaka:2005pn}, which introduces two almost degenerate HN with masses $M_{N1} \approx M_{N2} \sim 1 $GeV,
leading to a successful BAU and a third HN with mass $M_{N3} \sim $keV to be a natural candidate for DM.

Recently, NOs experiments have shown that the mixing-angle $\theta_{13}$ is non zero~\cite{An:2012eh} and also suggest 
the possibility of $CP$ violation in the light neutrino sector \cite{Abe:2018wpn}. However, extra sources of $CP$ violation are 
needed in order to explain the BAU via leptogenesis (see \cite{Chun:2017spz} for a review). In addition, when heavy neutrino 
masses are below the electroweak scale ($M_N < 246$ GeV), the BAU is generated via CP-violating heavy neutrino oscillations 
(HNOs) during their production \cite{Drewes:2016gmt}. 

In a previous article \cite{Zamora-Saa:2016ito} we have studied the resonant CP-violation and described the effects of 
HNOs on it. The study was carried out in the context of lepton number violating (LNV) tau lepton 
decay ($\tau^{\pm} \to \pi^{\pm} \pi^{\pm} \mu^{\mp}$) via two almost degenerate heavy on-shell 
Majorana neutrinos ($M_{N_i} \sim 1$GeV), which can oscillate among themselves. The purpose of this letter is to explore 
more realistic experimental conditions in order to observe such HNOs, extending the analysis beyond the resonant CP-violating scenario.

The work is arranged as follows: In Sec.~\ref{s3}, we study the production of the heavy neutrinos in tau's decays. In 
Sec.~\ref{sec:simu}, we present the results of the simulation of the HN production. In Sec.~\ref{sec:dis}, we present 
the results and shows conclusions.

%=================================================================================================================
\section{Production of the RHN}
\label{s3}
\begin{figure}[H]
\centering
\includegraphics[scale = 0.7]{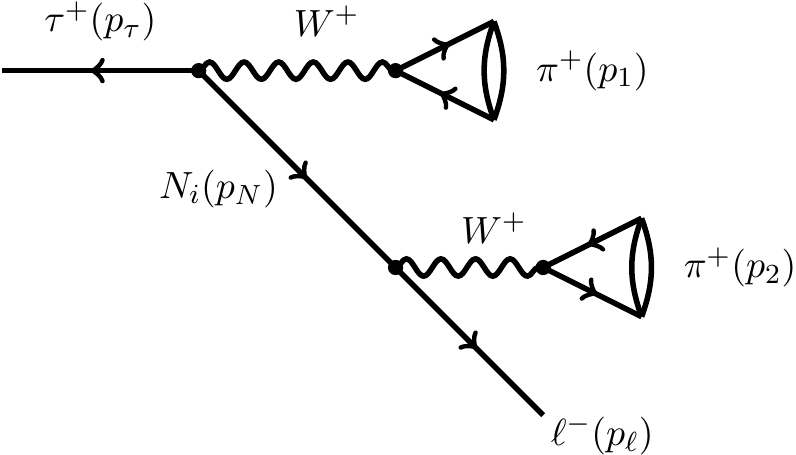}\hspace{0.3 cm}
\includegraphics[scale = 0.7]{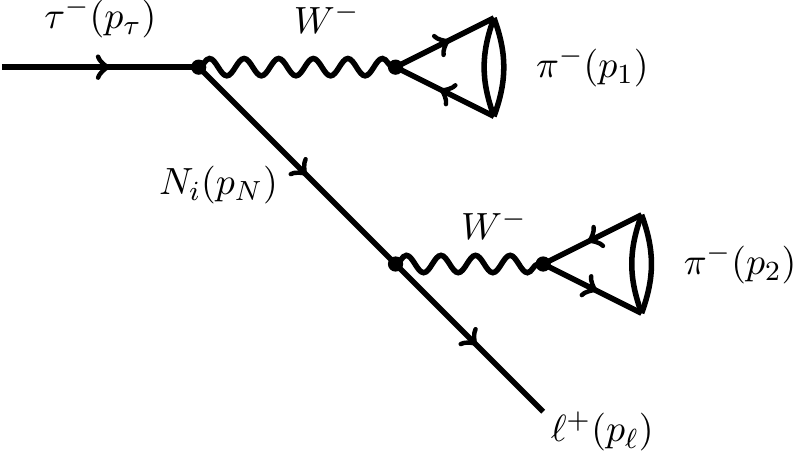}
\caption{Heavy neutrino production in tau lepton decay. Left Panel: Feynman diagrams for the LNV process $\tau^+ \rightarrow \pi^+ \pi^+ \ell^{-} $. Right Panel: Feynman diagrams for the LNV process $\tau^- \rightarrow \pi^- \pi^- \ell^{+}$}
\label{fig:taudecays}
\end{figure}

%==================================================================================================================
As established in the previous article \cite{Zamora-Saa:2016ito}, we are interested in studying the LNV processes which are represented by the Feynman diagrams shown in Fig.~\ref{fig:taudecays}, and from this point on, we will focus on the case of $\ell = \mu$. The heavy neutrinos $N_1$ and $N_2$ studied in this letter are almost degenerate ($M_{N1} \approx M_{N2}$) and  the mass difference\footnote{The neutrino ($N_{i}$) total decay width is expressed as $\Gamma_{Ni}$, the factor $\Gamma_{N}$ stand for $\Gamma_{N}=(\Gamma_{N1}+\Gamma_{N2})/2$ and $Y$ represent a parameter which allows us to express the mass difference in terms of $\Gamma_{N}$.} ($|\Delta M_N|=M_{N2}-M_{N1} \equiv Y \Gamma_{N} $) is in the range $5~\Gamma_{N} \leq |\Delta M_N| \leq 10~\Gamma_{N}$.
%%%%%%%%%%%%%%%%%%%%%%%%%%%%
%%%%%%%%%%%%%%%%%%%%%%%%%%%%%%

\begin{figure}[H]
\centering
\includegraphics[scale = 0.65]{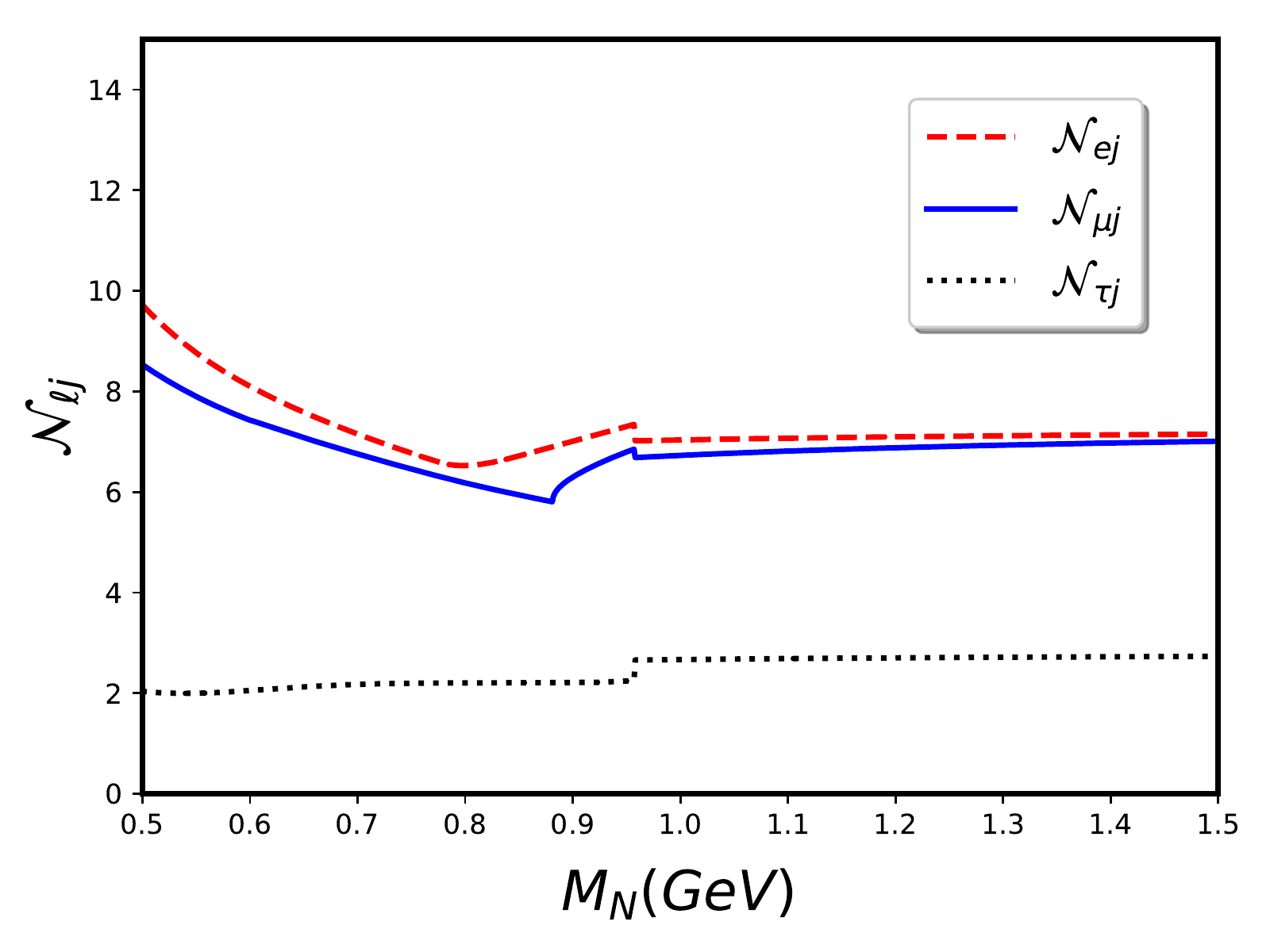}
\caption{Effective mixing coefficients ${\cal N}_{\ell j}^{\rm Ma}$ for Majorana neutrinos. Figure taken from \cite{Zamora-Saa:2016ito}.}
\label{fig:efcoef}
\end{figure}

\begin{figure}[H]
\centering
\includegraphics[scale = 0.49]{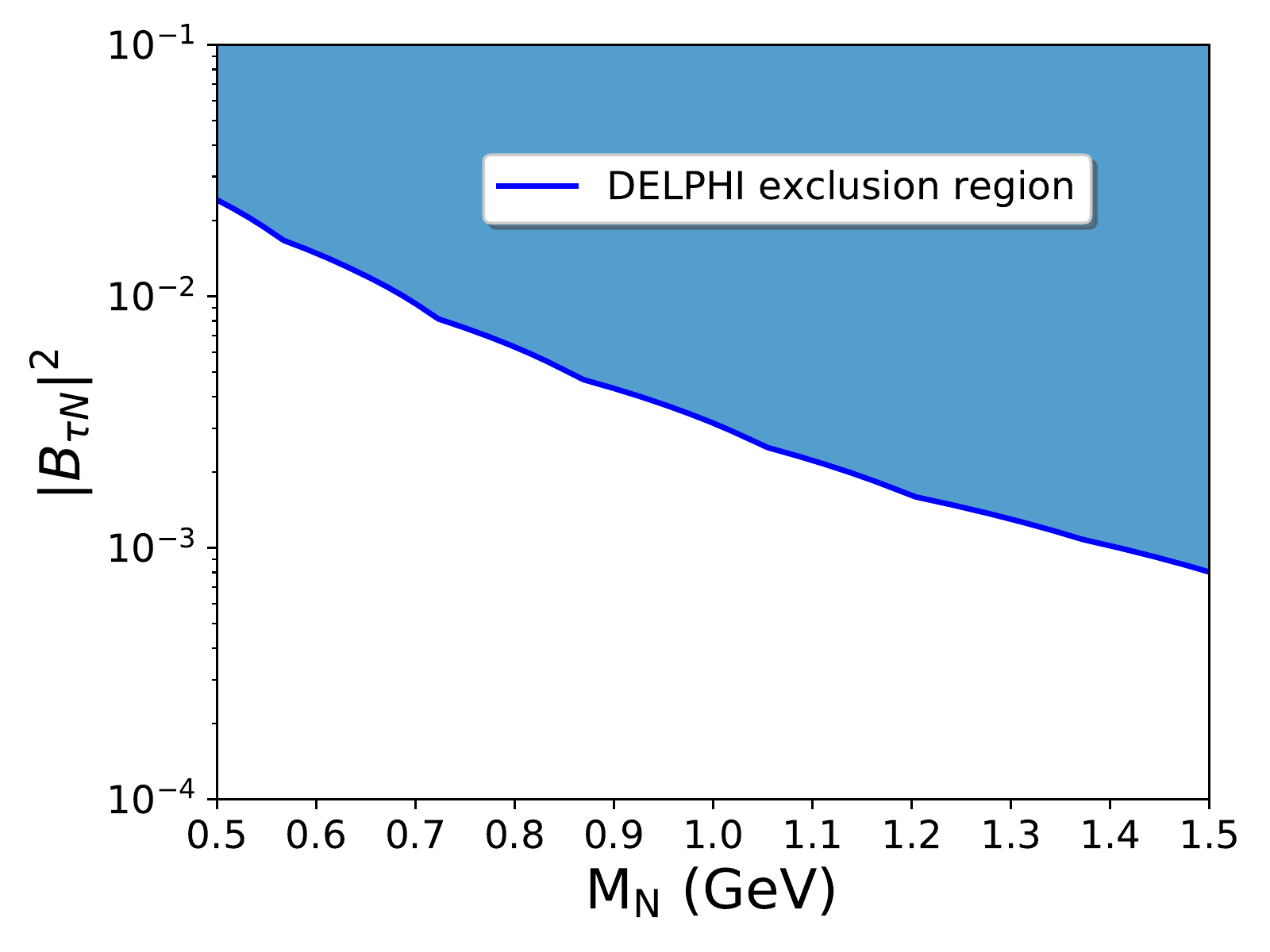}\hspace{0.1 cm}
\includegraphics[scale = 0.49]{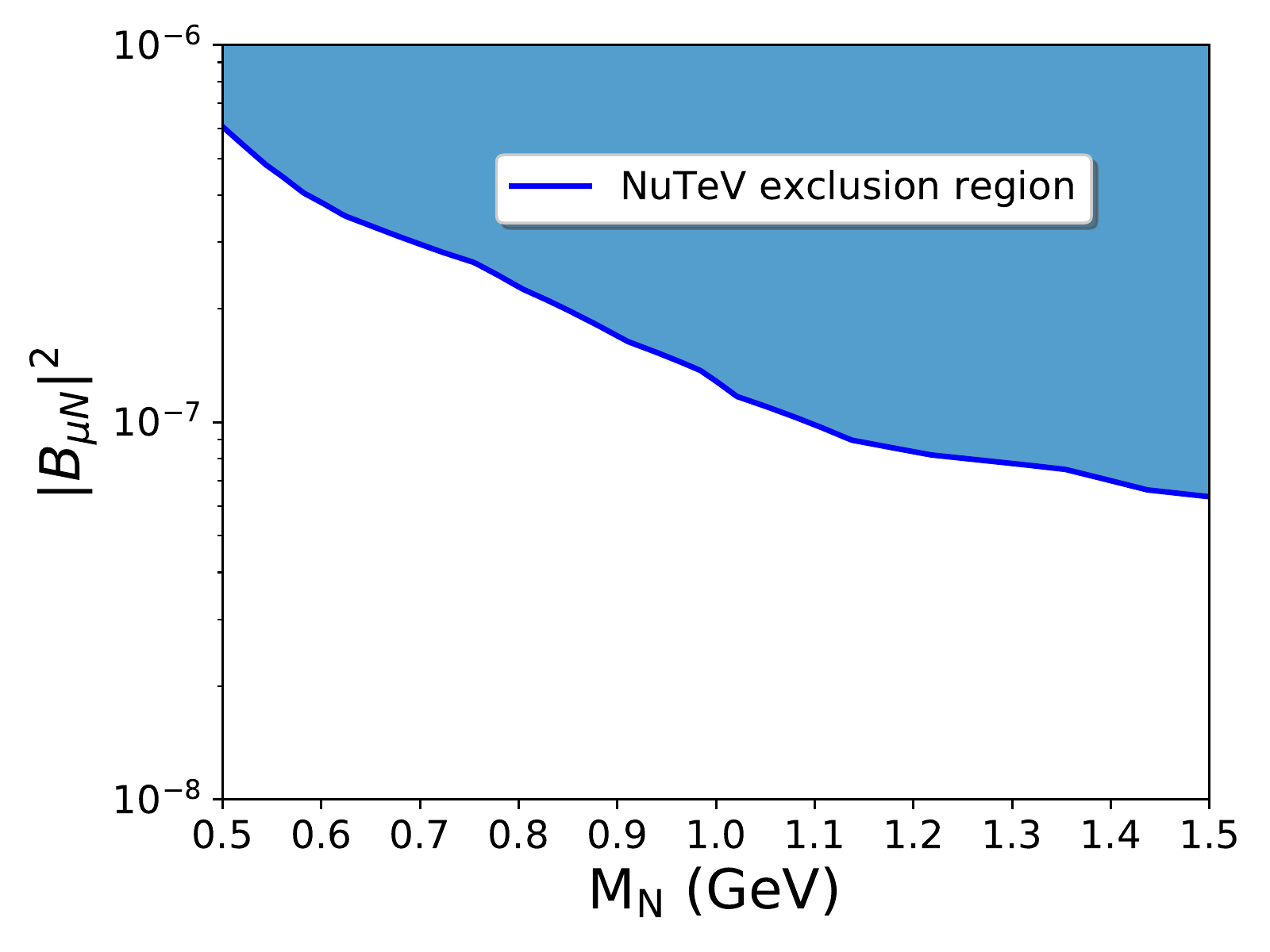}
\caption{Heavy neutrino mixing $|B_{\ell N}|^2$ exclusion regions. Data taken from Ref.~\cite{Atre:2009rg} and references from~\cite{Abreu:1996pa,Vaitaitis:1999wq}.}
\label{fig:mixlim}
\end{figure}

The relevant expressions for the aforementioned processes were presented in \cite{Cvetic:2015ura,Zamora-Saa:2016ito} as a function of the distance between production and detection vertices, called $L$. Therefore, the $L$ dependent effective differential decay width is given by
\begin{align}
\label{DWL}
\frac{d}{d L}& \Gamma_{\rm eff}^{\rm (osc)}(\tau^+ \to \pi^+ \pi^+ \mu^-;L) 
 \approx  \frac{1}{\gamma_N \beta_N}
\overline{\Gamma}(\tau^{+} \to \pi^{+} N) \overline{\Gamma}(N \to \pi^{+} \mu^{-})
\nonumber\\
& \times
\left\{ \sum_{j=1}^2 |B_{\mu N_j}|^2 |B_{\tau N_j}|^2 +
2  |B_{\mu N_1}| |B_{\tau N_1}| |B_{\mu N_2}| |B_{\tau N_2}|
\cos\left( L \frac{\Delta M_N}{ \gamma_N \beta_N} + \theta_{LV} \right)
\right\}.
\end{align}
Here, $|B_{\ell N_i}|$ are the mixing coeficients ($\ell = \mu ,\tau$ and $i=1,2$); the angle $\theta_{LV}$ stands for the CP-violating phase; the factors $\gamma_N \beta_N$ are the Lorentz factors and the heavy neutrino velocity\footnote{Ref.~\cite{Zamora-Saa:2016ito} considers $\gamma_N \beta_N$ of the produced $N_j$'s (in the laboratory frame) as fixed parameters $\gamma_N \beta_N=2$. However, the product $\gamma_N \beta_N$ is in general not fixed, because $\tau$ is moving in the lab frame when it decays into $N$ and $\pi$.}, respectively (see Appendix~I for more details). The factors $\overline{\Gamma}(\tau^{+} \to \pi^{+} N)$ and  $\overline{\Gamma}(N \to \pi^{+} \mu^{-})$ are the canonical partial decay widths (without mixing factors), which can be written as
\begin{subequations}
\label{bG}
\bea
\overline{\Gamma}(\tau^{\pm} \to \pi^{\pm} N) & = &
\frac{1}{8 \pi} G_F^2 f_{\pi}^2 |V_{ud}|^2 \frac{1}{M_{\tau}} 
\; \lambda^{1/2}\left(1, \frac{M_{\pi}^2}{M_{\tau}^2}, \frac{M_N^2}{M_{\tau}^2}\right)\times
\nonumber \\ 
&& \left[ \Big(M_{\tau}^2 - M_{N}^2\Big)^2 - M_{\pi}^2 \Big( M_{\tau}^2 + M_{N}^2 \Big) \right] \ ,
 \\
\overline{\Gamma}(N \to\pi^{\pm}  \ell^{\mp} ) & = &
\frac{1}{16 \pi} G_F^2 f_{\pi}^2 |V_{ud}|^2 \frac{1}{M_N} 
\; \lambda^{1/2}\left(1, \frac{M_{\pi}^2}{M_N^2}, \frac{M_{\ell}^2}{M_N^2}\right)\times 
\nonumber \\ 
&& \left[ \Big(M_N^2 + M_{\ell}^2\Big) \Big(M_N^2 - M_{\pi}^2+M_{\ell}^2\Big) -
4 M_N^2 M_{\ell}^2 \right]\ ,
\eea
\end{subequations}
where $G_F=1.166 \times 10^{-5}\ GeV^{-2}$ is the Fermi coupling constant,  $|V_{ud}|=0.974$ is a CKM matrix element and $f_{\pi}\approx 130.4\ MeV$ the pion decay constant.
The total heavy neutrino decay width $ \Gamma_{\rm Ma}(M_{N_i})$ is given by
\begin{equation}
  \Gamma_{\rm Ma}(M_{N_i}) \equiv \Gamma_{N_i}   \approx  \K_i^{\rm Ma}\ \frac{G_F^2 M_{N_i}^5}{96\pi^3} 
\label{DNwidth}
\end{equation}
 where  $\K_i^{\rm Ma}$ accounts for the mixings elements and reads as
 \begin{equation} 
 \K_i^{\rm Ma} = {\cal N}_{e i}^{\rm Ma} \; |B_{e N_i}|^2 + {\cal N}_{\mu i}^{\rm Ma} \; |B_{\mu N_i}|^2 + {\cal N}_{\tau i}^{\rm Ma} \; |B_{\tau N_i}|^2.
\label{DNwidth1}
\end{equation}
Here, ${\cal N}_{\ell i}^{\rm Ma}$ are the effective mixing coefficients, which account for all possible decay channels of $N_i$ (see Refs.~\cite{Atre:2009rg,Bondarenko:2018ptm}) and are presented in Fig.~\ref{fig:efcoef}. We note that for our mass range of interest ${\cal N}_{\ell i}^{\rm Ma} \sim 1$. In Eq.~\ref{DNwidth1}, the first two terms include both charged current and neutral current decays, whereas the third term arises purely due to neutral current decays. The neutral current decays are calculated in the approximation $|B_{\ell N_i}|^2 \ll 1$.
%%%%%%%%%%%%%%%%%%%%%%%%%%%%%%%%%%%%%%%%%%%%%%%%%%
%%%%%%%%%%%%%%%%%%%%%%%%%%%%%%%%%%%%%%%%%%%%%%%%%%
It is important to note that the mixings $|B_{\ell N_1}|^2$ and $|B_{\ell N_2}|^2$ can be different for the two heavy neutrinos, and consequently, the factors $\mathcal{K}^{\rm Ma}_i$ $(i = 1, 2)$ might be dissimilar from each other. However, in this letter we will assume that $|B_{\ell N_1}|^2 = |B_{\ell N_2}|^2$ and thus $\mathcal{K}^{\rm Ma}_1 \approx \mathcal{K}^{\rm Ma}_2$ $(\equiv \mathcal{K})$. In addition,  we  focus  on  the scenario in  which $|B_{\tau N}|$ mixing  parameter  is  much  larger  than  the  other  mixings (i.e. $|B_{\tau N_i}|^2 \gg |B_{\mu N_i}|^2 \sim |B_{e N_i}|^2$). We have chosen this scenario since methods for constraining $|B_{\tau N}|$ are lacking, so that it is much less constrained than $|B_{\mu N}|$ and $|B_{e N}|$, particularly in our mass range of interest $M_{N} < {m_{\tau}}$ (see Refs.~\cite{Atre:2009rg,Abreu:1996pa,Vaitaitis:1999wq} and references therein). Furthermore,  according to Fig.~\ref{fig:efcoef} we will assume that ${\cal N}_{\tau i}^{\rm Ma} \approx 2.5$ and ${\cal N}_{e i}^{\rm Ma} \approx {\cal N}_{\mu i}^{\rm Ma} \approx 7.5$. With these assumptions, we infer from Eq.~\eqref{DNwidth1} and Figs.~\ref{fig:efcoef}, \ref{fig:mixlim} that both heavy neutrinos have approximately the same total decay width. Additionally, since the $\tau$ channel\footnote{According 
$|B_{\tau N_i}|^2 \gg |B_{\mu N_i}|^2 \sim |B_{e N_i}|^2$ and ${\cal N}_{\ell i}^{\rm Ma} \sim 1 $ the factor $\K$ is approximated as $\K = {\cal N}_{e i}^{\rm Ma} \; |B_{e N_i}|^2 + {\cal N}_{\mu i}^{\rm Ma} \; |B_{\mu N_i}|^2 + {\cal N}_{\tau i}^{\rm Ma} \; |B_{\tau N_i}|^2 \approx {\cal N}_{\tau i}^{\rm Ma} \; |B_{\tau N_i}|^2$.} dominates $\Gamma_{\rm Ma}(M_{N_i})$ , we can write
\begin{equation}
\Gamma_{\rm Ma}(M_{N_i}) \equiv \Gamma_N(M_N) \approx 2.5 |B_{\tau N}|^2 \ \frac{G_F^2 M_{N}^5}{96\pi^3} \, .
\end{equation}
We note that CP violating phase ($\theta_{LV}$) can be extracted by means of the difference between the $L$-dependent effective differential decay width for $\tau^+$ and $\tau^-$
\begin{small}
\begin{align}
\nonumber
\frac{d}{dL} \Gamma(\tau^{+})-\frac{d}{dL} \Gamma(\tau^{-}) =& \frac{ -1}{\gamma_N \ \beta_N}  \; \overline{\Gamma}\big( \tau^+ \to \pi^+ N \big) \ \overline{\Gamma}\big( N \to \pi^+ \mu^-  \big)  \\
& \times 4 \ |B_{\mu N_1}| |B_{\tau N_1}| |B_{\mu N_2}| |B_{\tau N_2}| \sin \Big(L \; \frac{\Delta M_N}{\gamma_N \beta_N} \Big) \sin \Big(\theta_{LV} \Big)  \ .
\label{effassym}
\end{align}
\end{small}
%Here we remarks that $\gamma_N \beta_N$ distribution is the same for processes involving $\tau^+$ and $\tau^-$ (see Fig.~\ref{fig:lambda}). 

%=================================================================================================================
%=================================================================================================================
%=================================================================================================================
%=================================================================================================================
\section{Heavy neutrino simulations and results}
\label{sec:simu}

\begin{figure}[H]
\centering
\includegraphics[scale = 0.5]{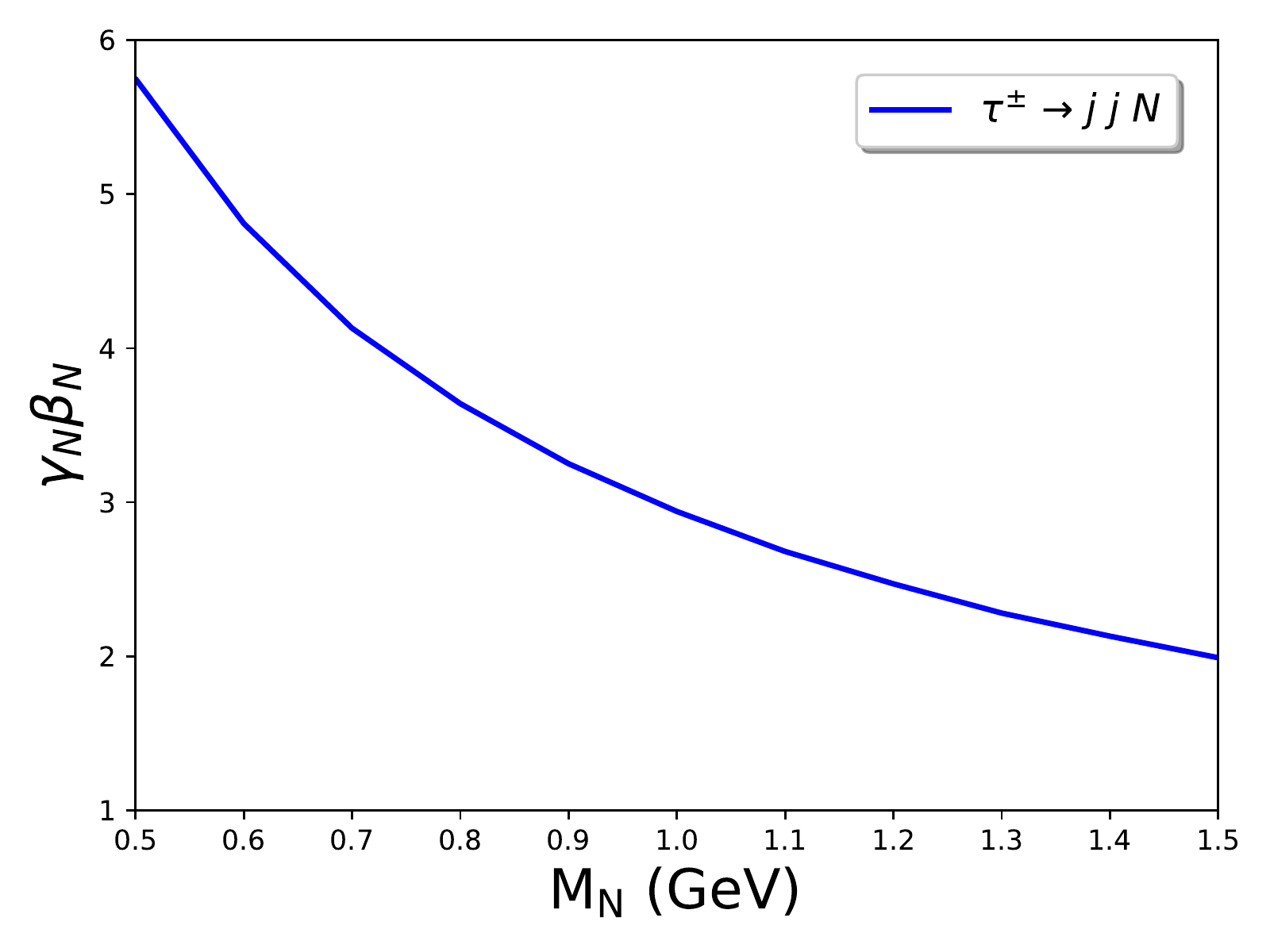}\hspace{0.01 cm}
\includegraphics[scale = 0.5]{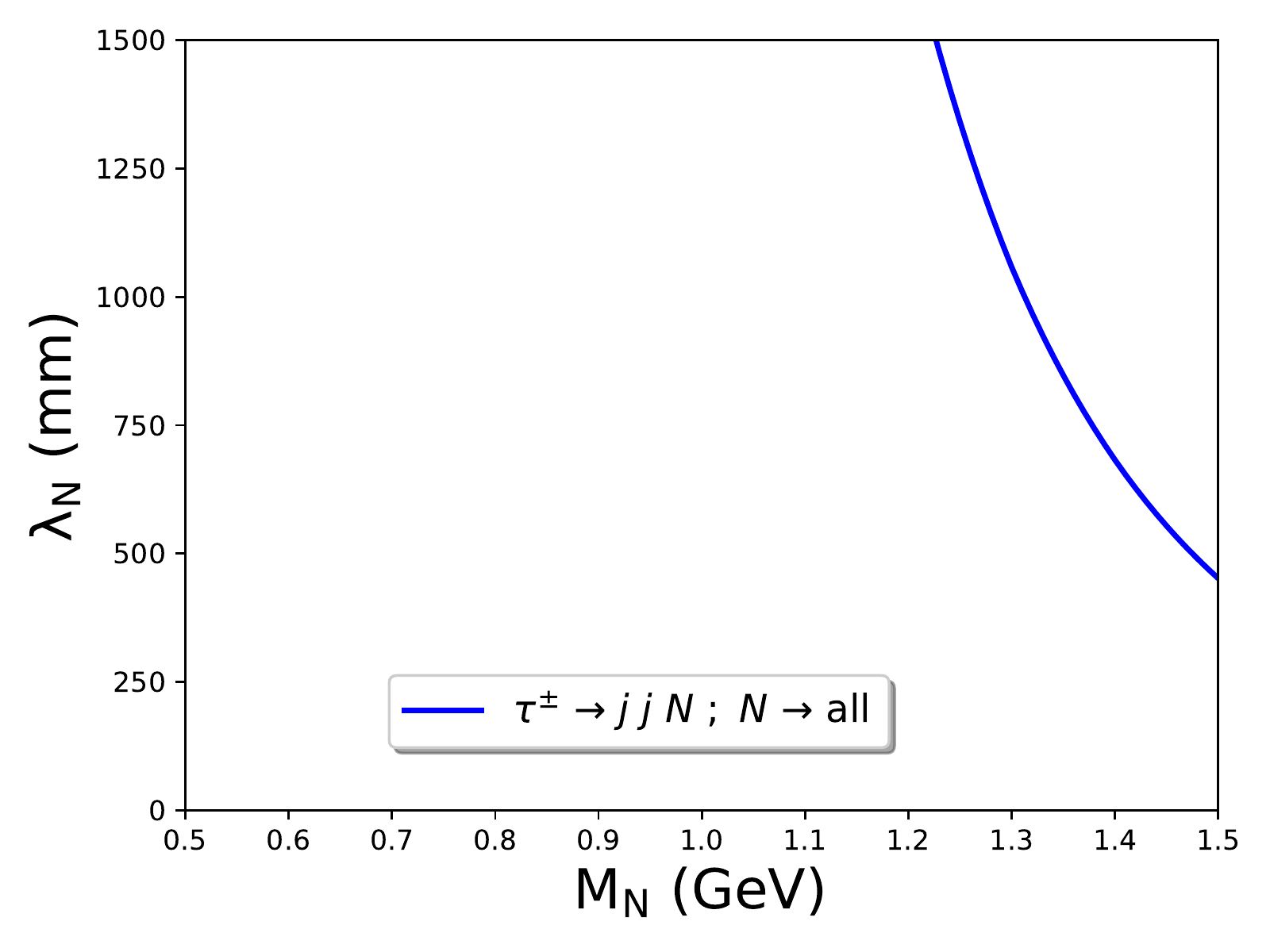}
\caption{{Left Panel}: Average heavy neutrino $\gamma_N \beta_N (= |{\vec p}_N|/M_N$) factor. {Right Panel}: Average heavy neutrino decay length $\lambda_N (= \gamma_N \beta_N/\Gamma_N$). Here we have taken $|B_{\tau N}|^2=10^{-3}$.}
\label{fig:lambda}
\end{figure}

We have simulated the $\tau^{\pm}$ production via the $e^+ e^- \to \tau^+ \tau^-$ process and its subsequent decay to HN ($e^+ e^- \to \tau^+ \tau^- \to q \bar{q} N$) in order to get a realistic $\gamma_{N}\beta_{N}$($\equiv|{\vec p}_N|/M_N$) distribution, here $q  \bar{q}$ stand for any light quark ($q=u,d,s$ and $\bar{q}=\bar{u},\bar{d}, \bar{s}$). We have carried out the simulation using \textsc{MadGraph5\_aMC@NLO}~\cite{Alwall:2014hca} for $\tau^{+}$ and $\tau^{-}$ individually, assuming Belle II kinematical parameters\footnote{The beam energies for $e^+$ and $e^-$ are $4$ GeV and $7$ GeV, respectively.}. The $\tau^+$ and $\tau^-$ do not show significant differences in their $\gamma_{N}\beta_{N}$ distributions, which are presented in the left panel of Fig.~\ref{fig:lambda}. The Universal FeynRules Output (UFO)~\cite{Degrande:2011ua} files were generated by means of FeynRules libraries~\cite{Alloul:2013bka}. 

It is important to point out that for our mass range of interest most of the heavy neutrinos $(N)$ tend to decay outside of the detector's radial acceptance $L_{D} \approx 1000$ mm (Fig.~\ref{fig:lambda} right-panel), introducing a strong suppression factor.
The factor $\gamma_{N}\beta_{N}$ is model by sampling a distribution obtained from the simulation (explained above) and used to evaluate $\frac{d \Gamma}{d L}(M_N,Y,\theta_{LV},L)$ as shown in Eq.~\ref{DWL}. For each value of $M_N$ ,$Y$ ,$ \theta_{LV}$ and $L$ Eq.~\ref{DWL} is sample with $10 000$ different values of $\gamma_{N}\beta_{N}$ to calculate the expected value of $\frac{d \Gamma}{d L}(M_N,Y,\theta_{LV},L)$.

%=================================================================================================================
 %=================================================================================================================
 %=================================================================================================================
 %=================================================================================================================
\section{Discussion of the results and summary}
\label{sec:dis}

In this work we have studied the modulation $d \Gamma/d L$ for the LNV process \\
$\tau^{\pm} \to \pi^{\pm} \pi^{\pm} \ell^{\mp}$ under Belle II conditions, 
in a scenario which contains two almost degenerate (on-shell) Majorana neutrinos $(N_j)$. This scenario has been studied in a previous work \cite{Zamora-Saa:2016ito} in which we explored the resonant CP-violation in rare tau decays. In that work, we found that when $Y=1$ the CP-violation is maximized and the heavy neutrino oscillation effects are negligible (NO HNO case). However, small deviations ($Y=5,10$) from $Y=1$ are allowed\footnote{In section 4 of the Ref.~\cite{Zamora-Saa:2016ito} this is calculated and explained in detail. } (HNO case) and may be relevant for explanations of BAU via leptogenesis \cite{Strumia:2006qk,Drewes:2016lqo,Drewes:2016gmt,Chun:2017spz}. 

\begin{figure}[t]
\centering
\includegraphics[width=0.49\textwidth]{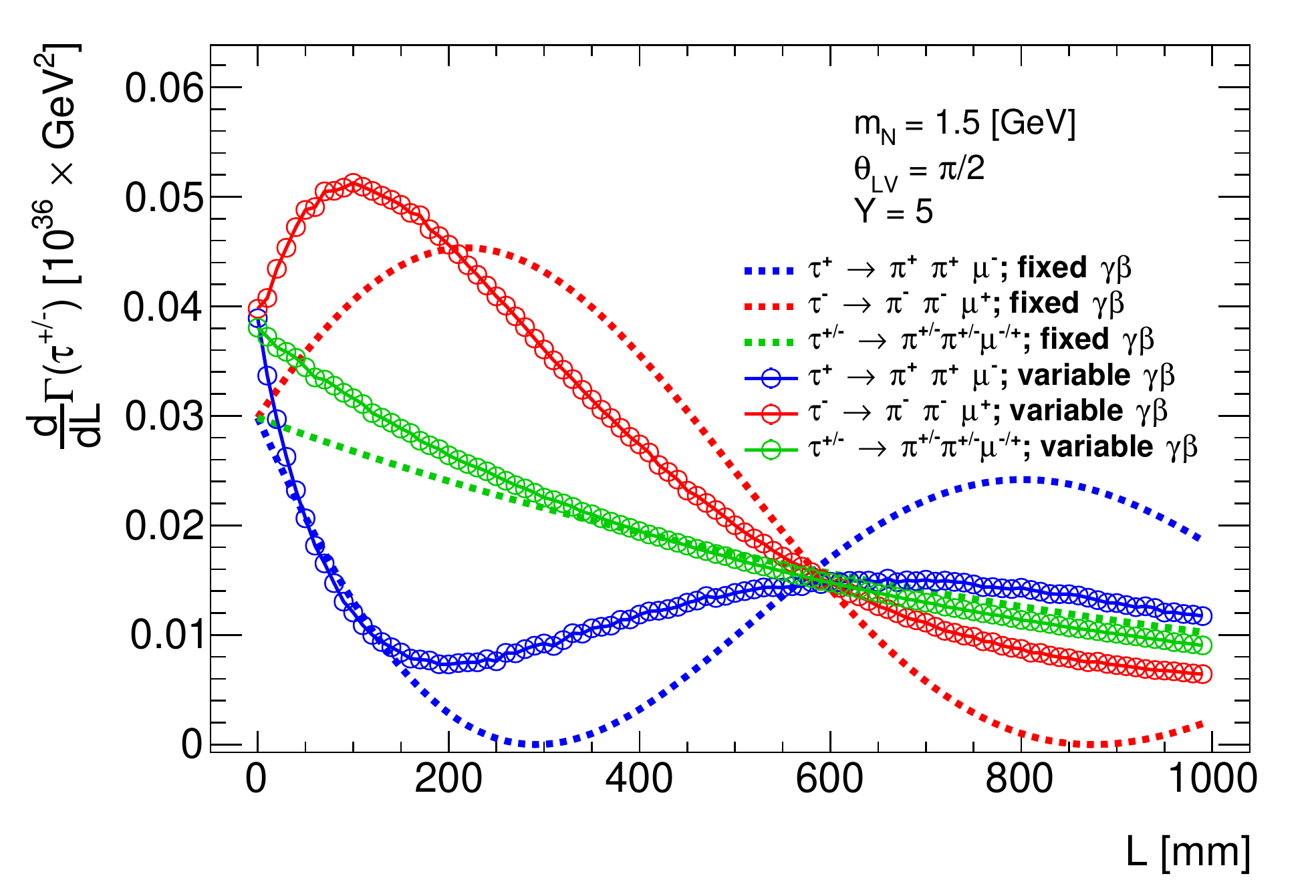}\hspace{0.01 cm}
\includegraphics[width=0.49\textwidth]{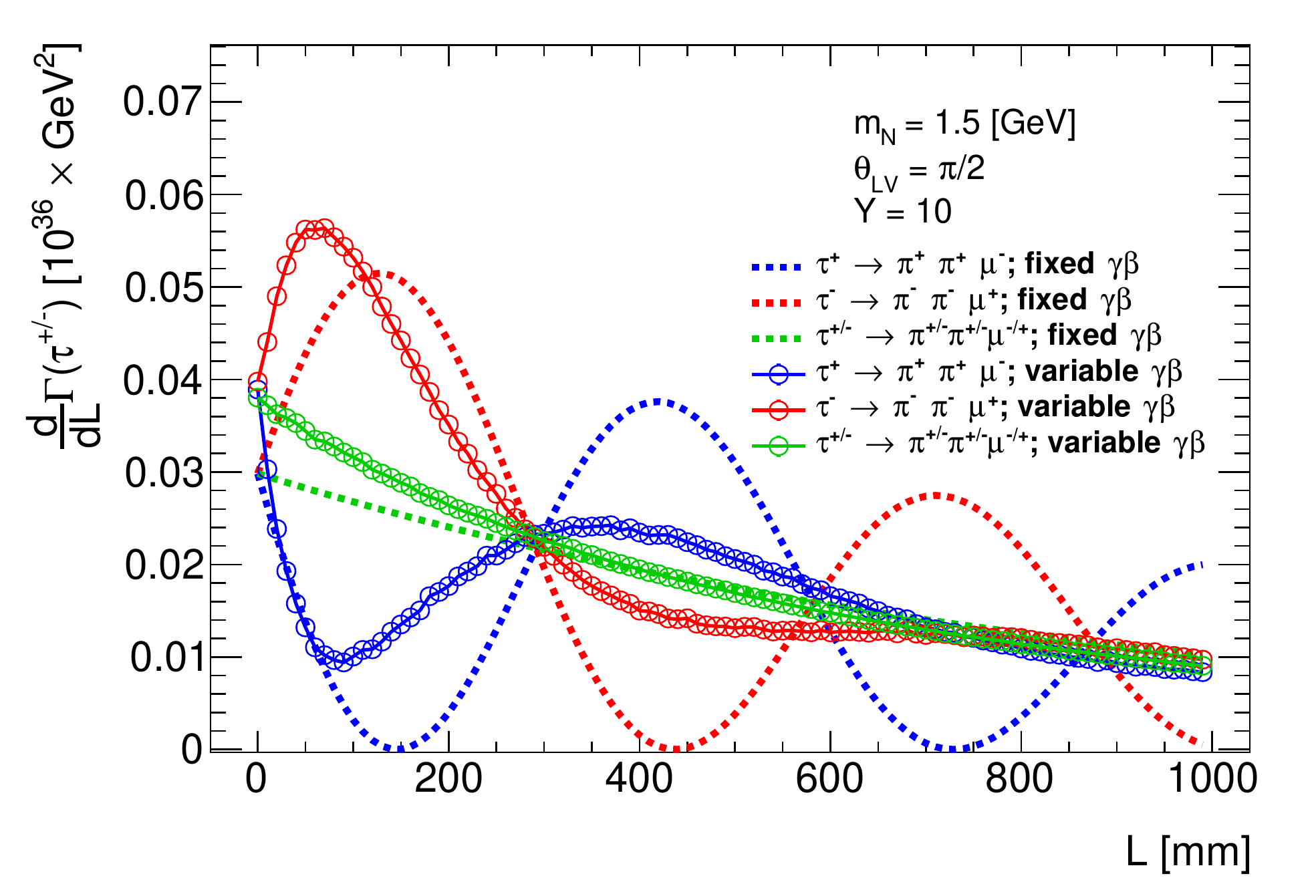}
\caption{Comparison of $d\Gamma/dL$ when we use a fixed value of $\gamma_{N}\beta_{N}$, and $d\Gamma/dL$ re-evaluated using the random sampling 
of $\gamma_{N}\beta_{N}$ from the simulation. Red and Blue colors stands for processes considering HNOs and the green one do not consider HNOs. Left panel: $M_N=1.5$ GeV, $\theta_{LV}=\pi/2$, $Y=5$. Right panel: $M_N=1.5$ GeV, 
$\theta_{LV}=\pi/2$, $Y=10$. Here we used $|B_{\tau N}|^2=5 \cdot 10^{-4}$ and $|B_{\mu N}|^2=5 \cdot 10^{-8}$.}
\label{fig:comparison}
\end{figure}

We note that the simulation of the production of on-shell heavy neutrinos, $N$, gave the same distribution of $\gamma_N \beta_N$ for both $\tau^+$ and $\tau^-$, and when it is considered, the modulation $d \Gamma/d L$ is smeared due to the fact that we have a distribution of small values of $\gamma_N \beta_N$ Fig.~\ref{fig:lambda} instead of a fixed (average) value (cf.~Fig.~\ref{fig:comparison}). In addition in studying the modulation for fixed and variable $\gamma_N \beta_N$ (Fig.~\ref{fig:comparison}) we can observe deviation of $d \Gamma / dL$ when HNOs are considered (blue and red lines) and when they are neglected (green lines).

\begin{figure}[h]
\centering
\includegraphics[width=0.49\textwidth]{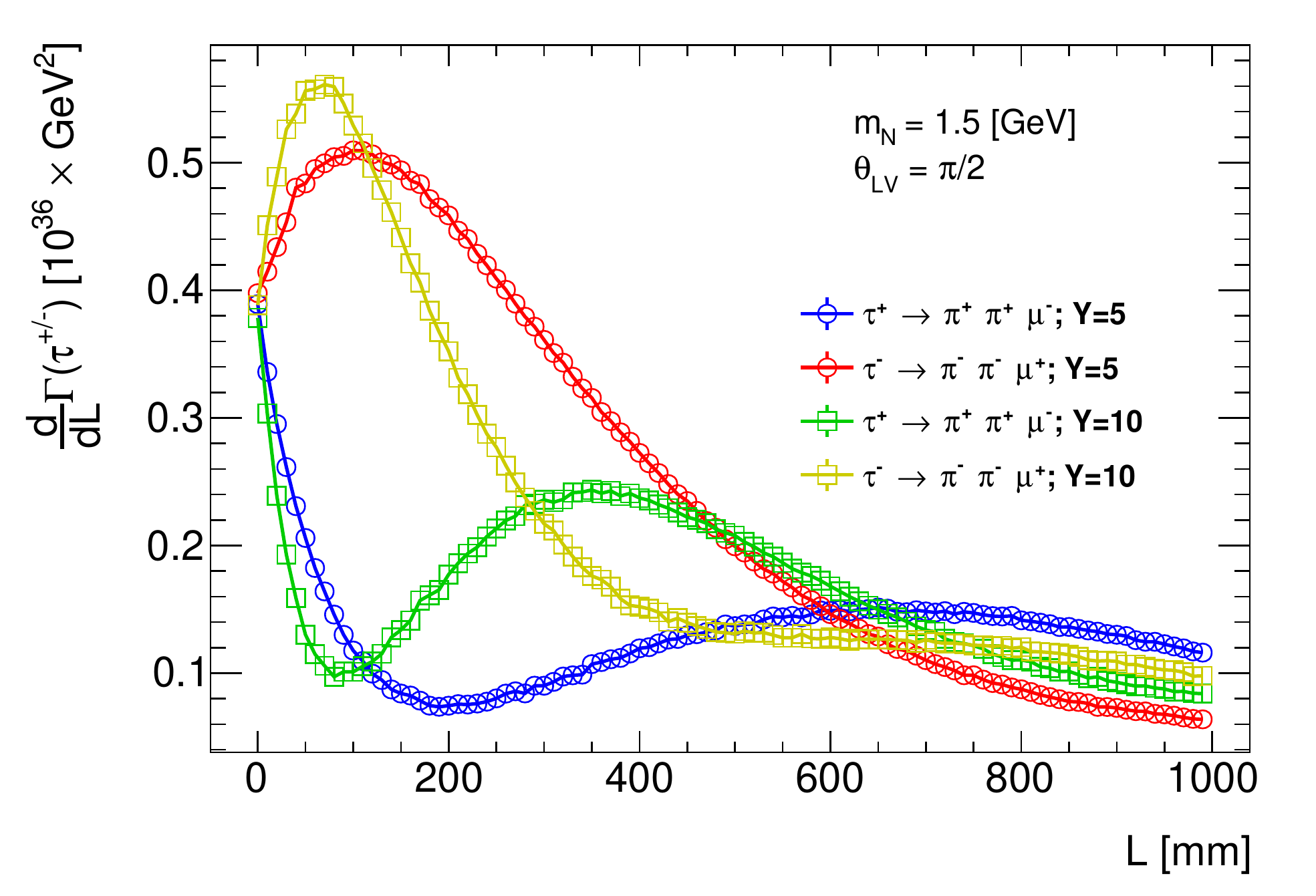}\hspace{0.01 cm}
\includegraphics[width=0.49\textwidth]{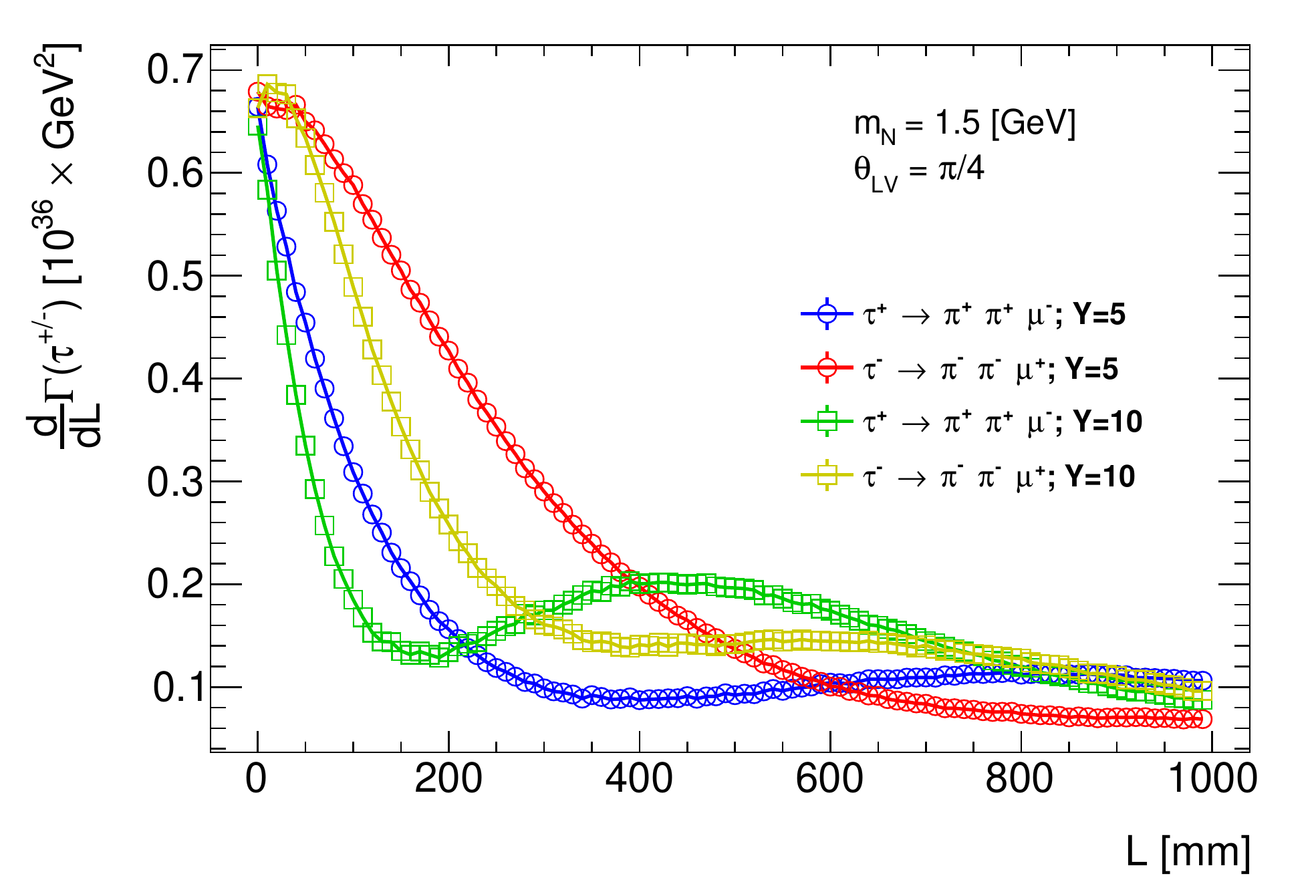}
\caption{Heavy neutrino oscillation modulation, with $\gamma \beta$ changing event-to-event. Left panel: $M_N=1.5$ GeV, $\theta_{LV}=\pi/2$ and $Y=5,10$. Right panel: $M_N=1.5$ GeV, $\theta_{LV}=\pi/4$ and $Y=5,10$. Here we used $|B_{\tau N}|^2=5 \cdot 10^{-4}$ and $|B_{\mu N}|^2= 5 \cdot 10^{-8}$.}
\label{fig:6}
\end{figure}

\begin{figure}[h]
\centering
\includegraphics[width=0.49\textwidth]{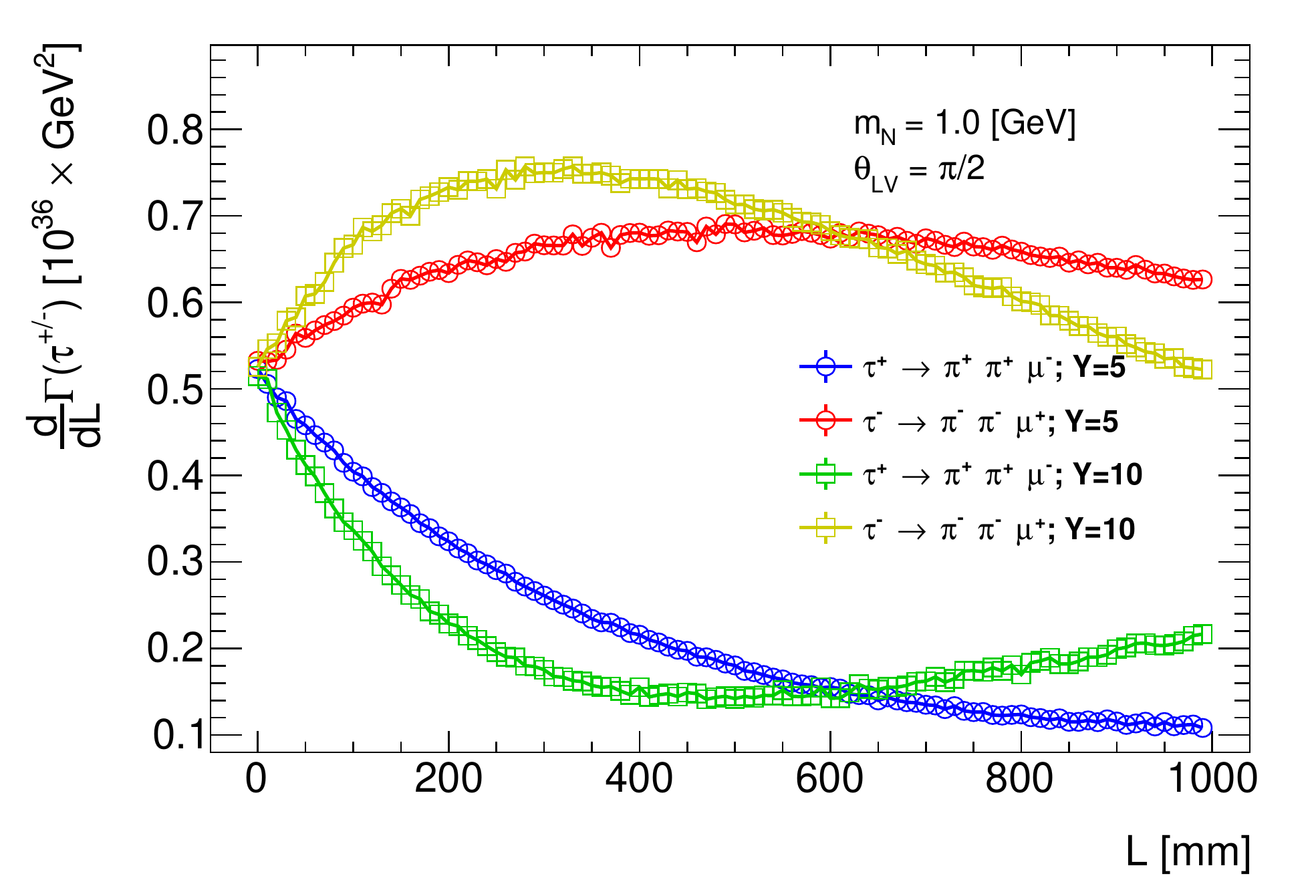}\hspace{0.01 cm}
\includegraphics[width=0.49\textwidth]{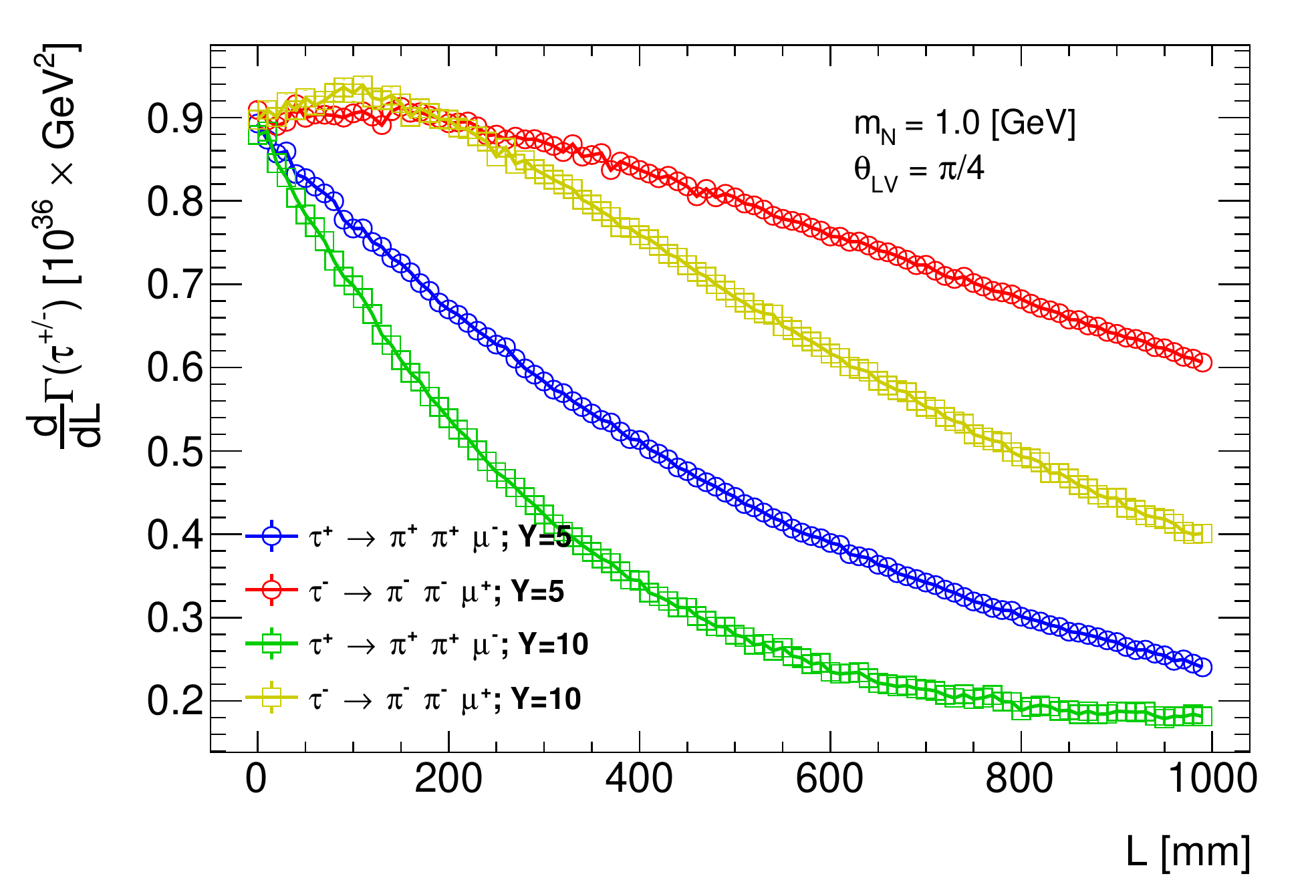}
\caption{Heavy neutrino oscillation modulation, with $\gamma \beta$ changing event-to-event. Left panel: $M_N=1.0$ GeV, $\theta_{LV}=\pi/2$ and $Y=5,10$. Right panel: $M_N=1.0$ GeV, $\theta_{LV}=\pi/4$ and $Y=5,10$. Here we used $|B_{\tau N}|^2=10^{-3}$ and $|B_{\mu N}|^2=10^{-7}$.}
\label{fig:7}
\end{figure}

\begin{figure}[h]
\centering
\includegraphics[width=0.49\textwidth]{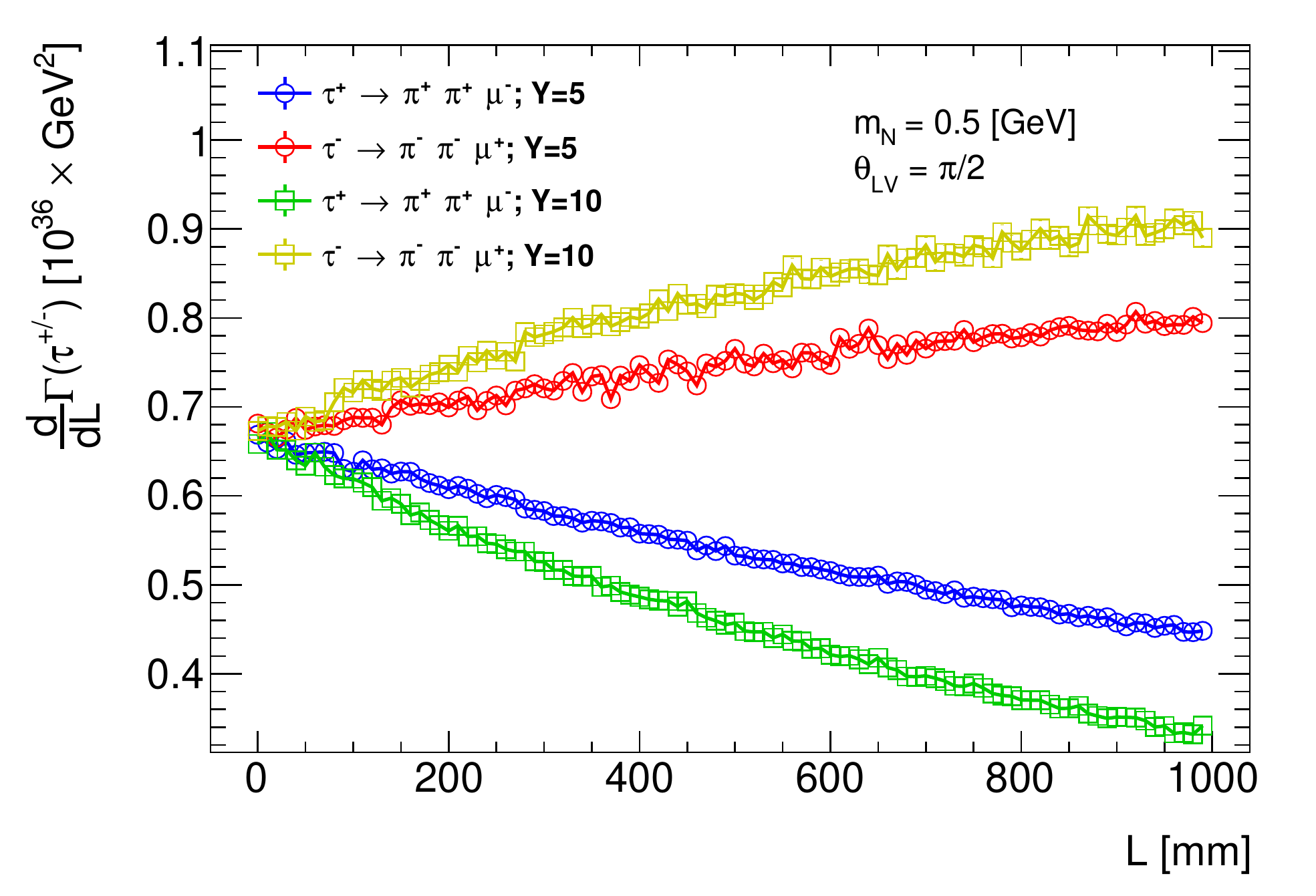}\hspace{0.01 cm}
\includegraphics[width=0.49\textwidth]{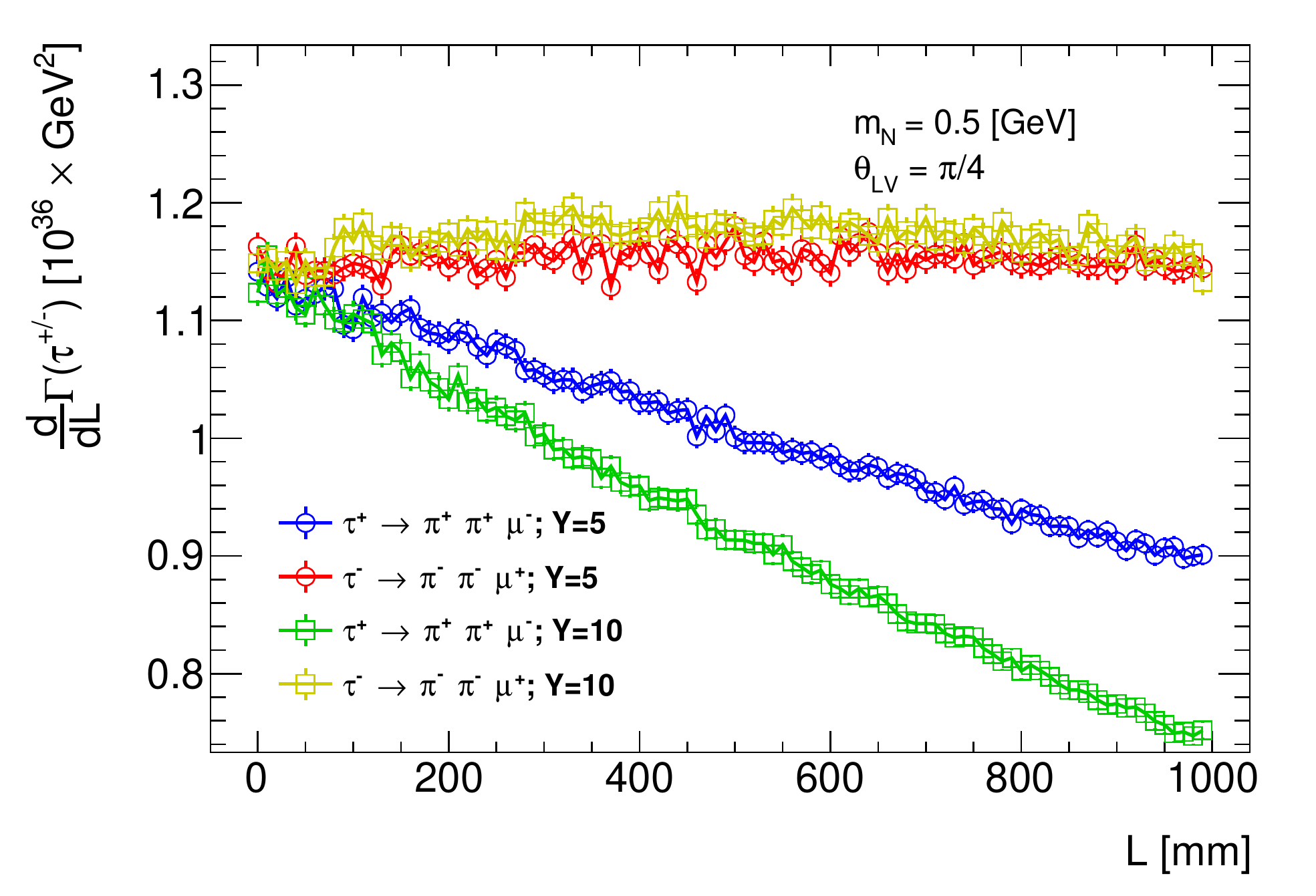}
\caption{Heavy neutrino oscillation modulation, with $\gamma \beta$ changing event-to-event. Left panel: $M_N=0.5$ GeV, $\theta_{LV}=\pi/2$ and $Y=5,10$. Right panel: $M_N=0.5$ GeV, $\theta_{LV}=\pi/4$ and $Y=5,10$. Here we used $|B_{\tau N}|^2=10^{-2}$ and $|B_{\mu N}|^2=10^{-7}$.}
\label{fig:8}
\end{figure}

We have studied the modulation $d \Gamma/d L$ for $\tau^{\pm}$ decays and for different values of the parameters $M_N$, $Y$, and the CP violating phase $\theta_{LV}$. In addition, we remark that in Figs.~\ref{fig:comparison} - \ref{fig:8} the number of events considered were almost infinite and the vertex resolution considered was $0.03$ mm \cite{Abe:2010gxa}. We find the modulation shape strongly dependent on the  CP-violation phase $\theta_{LV}$, which supports the possibility of obtaining the value of $\theta_{LV}$ from measurements of $d \Gamma/d L$.

In Fig.~\ref{fig:6} (left panel), when $M_N=1.5$ GeV and $\theta_{LV}=\pi/2$, we observe that the number of expected $\tau^+$ decays inside the region $100 \leq L \leq 650$ mm is larger for $Y=10$ than $Y=5$; with difference in the remaining regions being negligible. On the other hand, when $M_N=1.5$ GeV and $\theta_{LV}=\pi/4$, we observe from Fig.~\ref{fig:6} (right panel) that for $\tau^+$ decays inside the region $50 \leq L \leq 200$ mm the number of expected events is larger for $Y=5$ than $Y=10$. Conversely, inside the region $200 \leq L \leq 800$ mm,  $Y=10$ dominates over $Y=5$, while for $L \geq 800$ mm the differences are negligible. For the $\tau^-$ decays, the situation changes drastically: for $M_N=1.5$ GeV and $\theta_{LV}=\pi/2$, the number of expected $\tau^-$ decays inside the region $50 \leq L \leq 100$ mm is larger for $Y=10$ than $Y=5$. Conversely inside $100 \leq L \leq 650$ mm $Y=5$ dominates over $Y=10$. When $M_N=1.5$ GeV and $\theta_{LV}=\pi/4$ inside the region $50 \leq L \leq 500$ mm, $Y=5$ dominates over $Y=10$, and for $500 \leq L \leq 1000$~mm, the opposite is true. 

In Fig.~\ref{fig:7} (left panel) when $M_N=1.0$ GeV and $\theta_{LV}=\pi/2$, we observed that the number of expected $\tau^+$ decays in the region $0 \leq L \leq 600$ mm are larger for $Y=5$ than $Y=10$. 
Conversely, from $600 \leq L \leq 1000$ mm, $Y=10$ dominates over $Y=5$. Futhermore, when $M_N=1.0$ GeV and $\theta_{LV}=\pi/4$, we observed from Fig.~\ref{fig:7} (right panel) that the number of expected $\tau^+$ decays in the entire region $0 \leq L \leq 1000$ mm is larger for $Y=5$ than $Y=10$. For the $\tau^-$ decays, the situation is different: for $M_N=1.0$ GeV and $\theta_{LV}=\pi/2$, the number of expected $\tau^-$ decays inside the region $0 \leq L \leq 600$ mm is larger for $Y=10$ than $Y=5$. Conversely, from $600 \leq L \leq 1000$ mm, $Y=5$ dominates over $Y=10$. When $M_N=1.0$ GeV and $\theta_{LV}=\pi/4$, in the region $200 \leq L \leq 1000$ mm $Y=5$ dominates over $Y=10$. In other regions the differences are negligible. 

In Fig.~\ref{fig:8} (left panel) when $M_N=0.5$ GeV and $\theta_{LV}=\pi/2$, we observed that the expected number of $\tau^+$ decays inside the whole region $0 \leq L \leq 1000$ mm are larger for $Y=5$ than $Y=10$. The same can be observed when $M_N=0.5$ GeV and $\theta_{LV}=\pi/4$ (Fig.~\ref{fig:8}, right panel). For the $\tau^-$ decays, the situation is different: for $M_N=0.5$ GeV and $\theta_{LV}=\pi/2$, the number of expected $\tau^-$ decays inside the whole region $0 \leq L \leq 1000$ mm is larger for $Y=10$ than $Y=5$. On the other hand, when $M_N=0.5$ GeV and $\theta_{LV}=\pi/4$, we observed from Fig.~\ref{fig:8} (right panel), that the difference between $Y=5$ and $Y=10$ is negligible for $\tau^-$ decays in the full range of $0 \leq L \leq 1000$ mm.

\begin{figure}[b]
\centering
\includegraphics[width=0.49\textwidth]{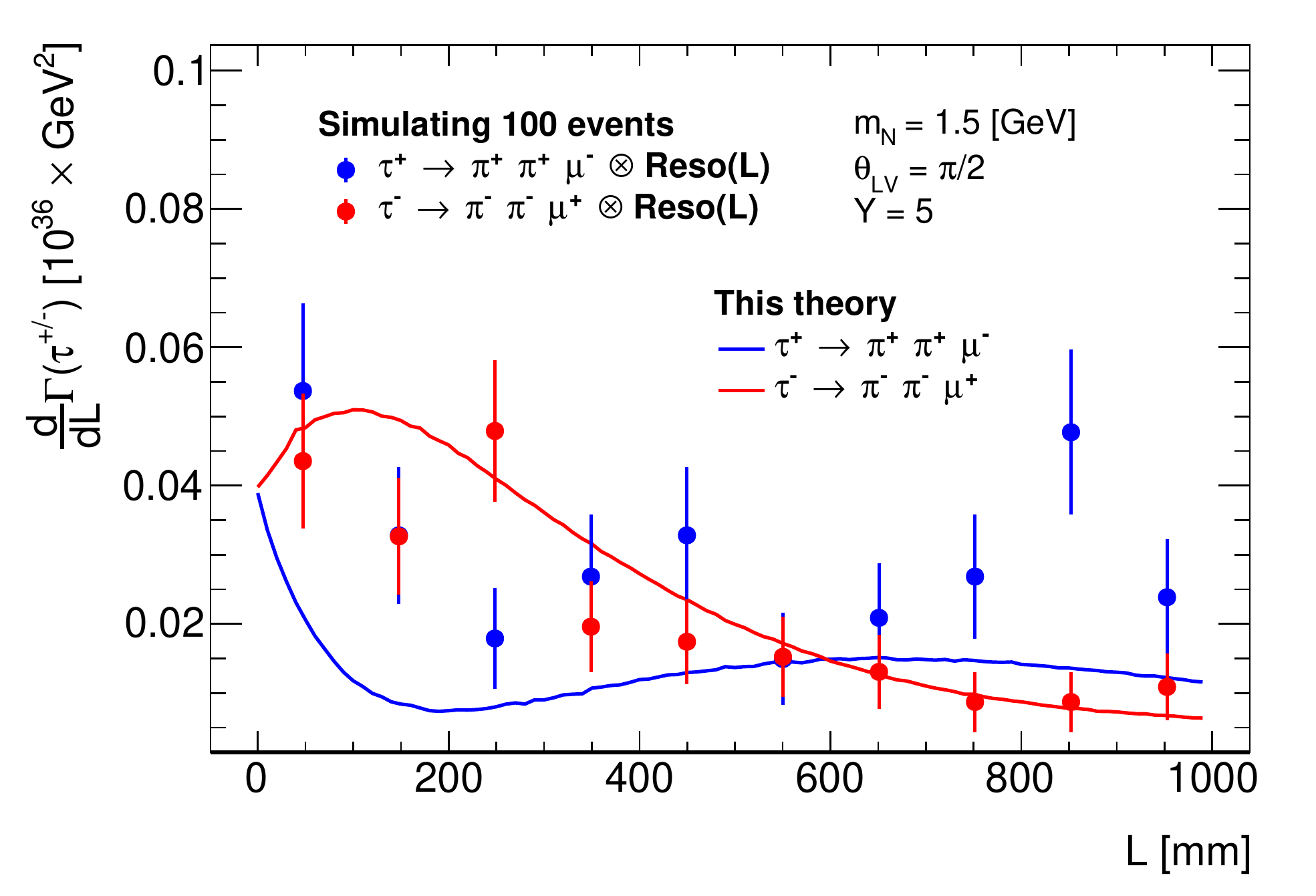}\hspace{0.01 cm}
\includegraphics[width=0.49\textwidth]{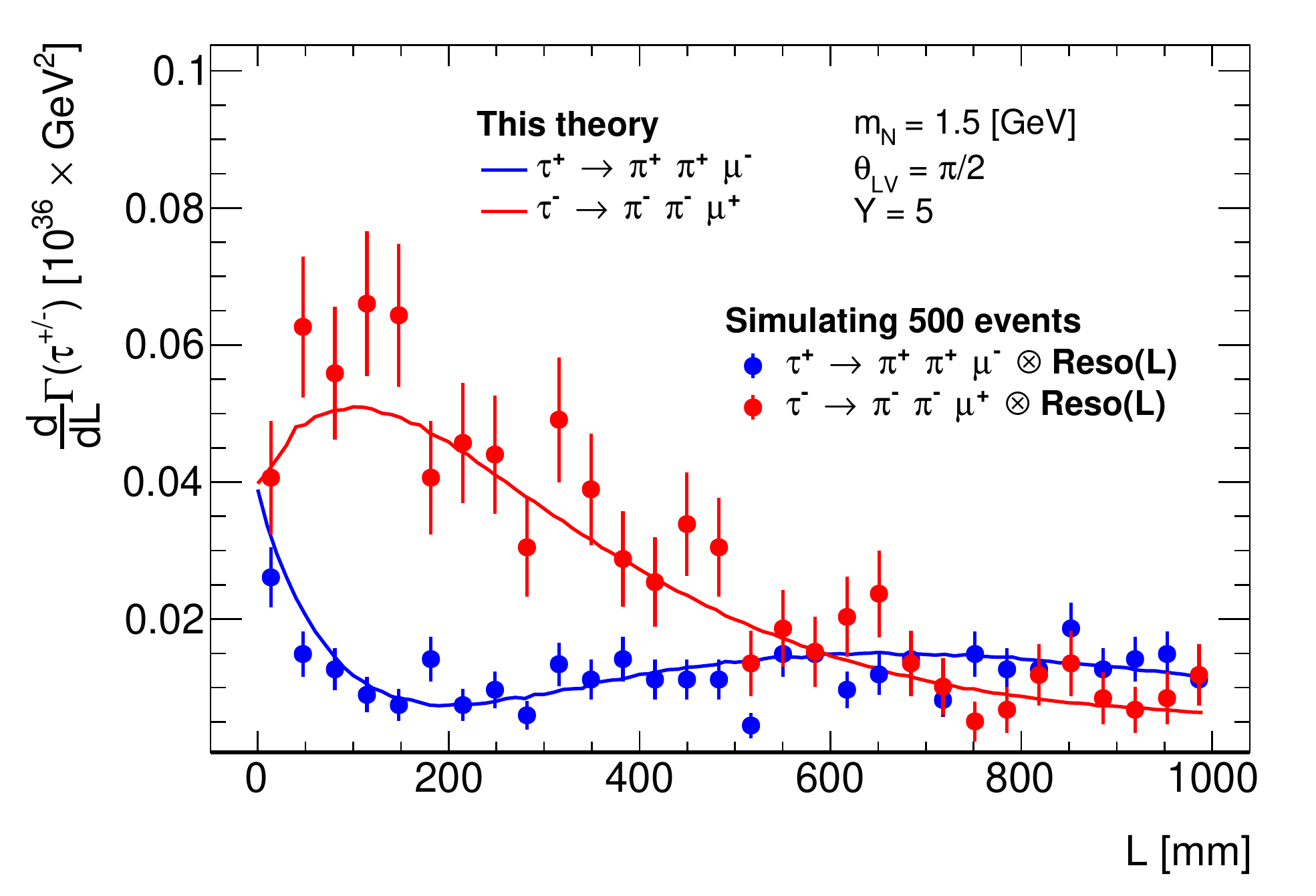}
\caption{The solid line stands for the heavy neutrino oscillation modulation, while the datapoints stand for random-samplings of the heavy neutrino oscillation modulation convolved with the detector resolution (${\rm Reso(L)}=0.03$ mm). Left panel: $M_N = 1.5$ GeV, $\theta_{LV} = \pi/2$, $Y = 5$ for 100 simulated events. Right panel: $M_N = 1.5$ GeV, $\theta_{LV} = \pi/2$, $Y = 5$ for 500 simulated events. Here we used $|B_{\tau N}|^2=5 \cdot 10^{-4}$ and $|B_{\mu N}|^2= 5 \cdot 10^{-8}$.}
\label{fig:9}
\end{figure}

In Fig.~\ref{fig:9}, we present results for a finite number of detected events including the statistical uncertainties, when $M_N=1.5$ GeV; $Y=5$ and $\theta_{LV}=\pi/2$. In the left panel, we present results for 100 simulated events and in the right panel, for 500 simulated events. Furthermore, the considered vertex-position resolution was $0.03$ mm \cite{Abe:2010gxa}. We notice for the case of 100 simulated events, the two distributions are quite similar. Which may preclude distinguishing them in experiment. However, in the case of 500 simulated events, we have enough statistical significance to separe the $\tau^+$ and $\tau^-$ modulation, in the range $50 \leq L \leq 500$ mm.

\newpage

In summary, in this work we have considered the heavy neutrino oscillations of $\tau^{\pm}$ decays in a scenario which contains two heavy, almost-degenerate neutrinos $(N_j)$, with masses in the range $0.5$ GeV $\leq M_N \leq 1.5$ GeV. We have explored the feasibility to measure CP-violating HNOs processes in such a scenario where the modulation of $d \Gamma/d L$ for the process $\tau^{\pm} \to \pi^{\pm} N \to \pi^{\pm} \pi^{\pm} \mu^{\mp}$ at Belle II can be resolved inside the detector. We have established some realistic conditions for  $|B_{\tau N}|^2$, $|B_{\mu N}|^2$ and $Y (\equiv \Delta M_N/\Gamma_N)$ where the aforementioned effect can be observed.

%=================================================================================================================
 %=================================================================================================================
 %=================================================================================================================
 %=================================================================================================================
\section{Acknowledgments}
%=============================================
This work is supported in part by FONDECYT Grant No.~3180032 (J.Z.S.). The work of S.T.A. is supported by the National Science Foundation (NSF) grant 1812377.
%=================================================================================================================
 %=================================================================================================================
 %=================================================================================================================
 %===========================================================================================================

\section*{Appendix I}
In Ref.~\cite{Zamora-Saa:2016ito} we have considered the parameters $\beta_N$ (HN velocity) and $\gamma_N$ (HN Lorentz factor) of the produced heavy neutrinos N in the laboratory frame ($\Sigma$) as fixed values ($\beta_N \gamma_N=2$). However, the factor $\beta_N \gamma_N$ is in general not fixed, due to the $\tau$ lepton is moving in the lab frame when it decays into $N$'s and $\pi$'s. The factor $\beta_N \gamma_N$ can be written as follow
\begin{equation}
\beta_N \gamma_N = \sqrt{(E_N({\hat p}'_N)/M_N)^2 - 1},
\label{bNgN}
\end{equation}
where the energy of the HN in the lab frame, $E_N$, depends on its direction ${\hat p}'_N$ in the $\tau$-rest frame ($\Sigma'$).
%%%%%%%%%%%%%%%%%%%%%%%%%%%%%%%%%%%%%%%
\begin{figure}[H]
\centering
\includegraphics[width=8cm]{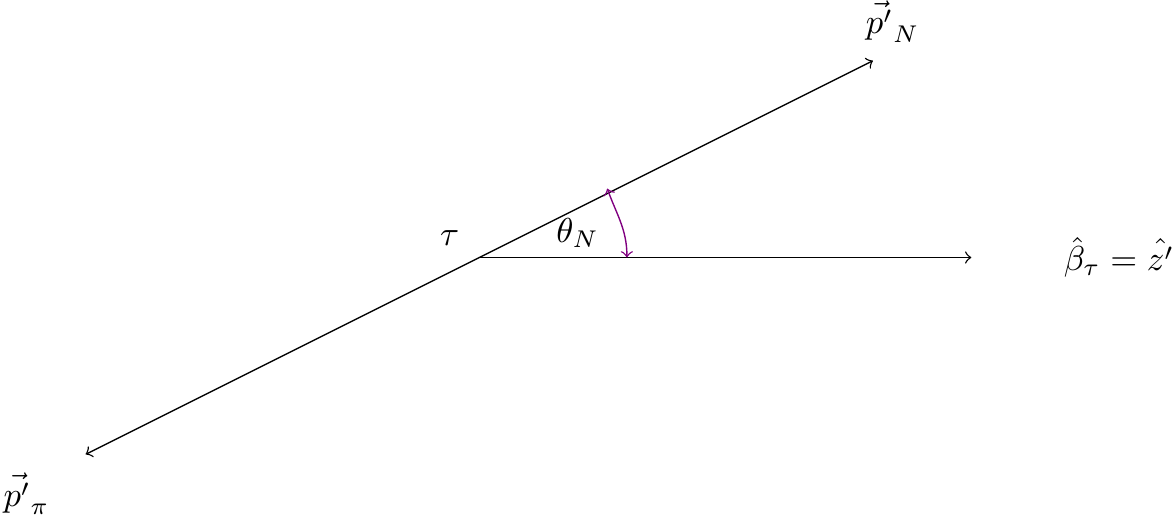}%\hspace{0.1cm}
\caption{Schematic particles 3-momentum representation in the $\tau$-rest frame ($\Sigma'$). The direction ${\hat \beta}_{\tau}$ of the velocity of $\tau$ in the lab frame defines the ${\hat z}'$-axis and $\theta_N$ stand for the angle between ${\hat \beta}_{\tau}$ and HN momentum ${\hat p}'_N$.}
 \label{FigpN}
\end{figure}
%%%%%%%%%%%%%%%%%%%%%%%%%%%%%%%%%%%%%%%%%
The HN energy $E_N$ can be written in terms of the angle $\theta_N$  and momentum ${\hat p}'_N$ as follow
\begin{equation}
E_N = \gamma_{\tau} (E'_N + \cos \theta_N \beta_{\tau} |{\vec p}'_N|),
\label{EN}
\end{equation}
where the corresponding quantities in the $\tau$-rest frame ($\Sigma'$) are fixed
\begin{equation}
E'_N = \frac{M_{\tau}^2 + M_N^2 - M_{\pi}^2}{2 M_{\tau}}, \quad
|{\vec p}'_N| = \frac{1}{2} M_{\tau} \lambda^{1/2} \left( 1, \frac{M_{\pi}^2}{M_{\tau}^2}, \frac{M_N^2}{M_{\tau}^2} \right),
\label{ENppNp}
\end{equation}
$\beta_{\tau}$ is the velocity of $\tau$ in the lab frame, and $\lambda$ is given by
\begin{equation}
\lambda(a,b,c)=a^2+b^2+c^2-2ab-2ac-2cb \ .
\end{equation}

Therefore, the Eq.~\ref{DWL} can be written as
\begin{small}
\begin{align}
  \nonumber \frac{d}{d L}& \Gamma_{\rm eff}^{\rm (osc)}(\tau^+ \to \pi^+ \pi^+ \mu^-;L)   \approx
\int  \frac{d \Omega_{{\hat p}'_N}}{\left[ (E_N({\hat p}'_N)/M_N)^2 - 1 \right]^{1/2} }  \; \frac{d \overline{\Gamma}\big( \tau^+ \to \pi^+ N \big)}{ d \Omega_{{\hat p}'_N}} \ \overline{\Gamma}\big( N \to \pi^+ \mu^-  \big)  \\
& \times \Bigg( \sum_{i=1}^2 |B_{\mu N_i}|^2 |B_{\tau N_i}|^2+ 2 |B_{\mu N_1}| |B_{\tau N_1}| |B_{\mu N_2}| |B_{\tau N_2}| \cos \Big(2\pi \; \frac{L}{L_{\rm osc}({\hat p}'_N)} + \theta_{LV} \Big)\Bigg) \ , 
\label{effdwfosc2}
\end{align}
\end{small}
where now the oscillation length $L_{\rm osc}$, appearing in the last term, also depends on the direction ${\hat p}'_N$
\begin{equation}
L_{\rm osc}({\hat p}'_N) = \frac{2 \pi\beta_N \gamma_N}{M_N} = 
\frac{2 \pi}{M_N^2} |{\vec p}_N({\hat p}'_N)| =
\frac{2 \pi}{M_N} \left[ (E_N({\hat p}'_N)/M_N)^2 - 1 \right]^{1/2}.
\label{Losc2}
\end{equation}

It is important to remarks that in a real experiment the produced $\tau$ leptons can have a wide range of momenta, these momenta can be well described by a distribution, which was simulated and obtained in the present work.

\bibliographystyle{apsrev4-1}

\bibliography{biblio.bib}

\end{document}